\documentstyle[aps,prb,twocolumn,epsf]{revtex}
\begin{document}
\input{psfig.tex} 

\title{Dynamical effects of the nanometer-sized polarized domains in
  Pb(Zn$_{1/3}$Nb$_{2/3}$)O$_3$ }

\author{P.\ M.\ Gehring} 

\address{NIST Center for Neutron Research, National Institute of
  Standards and Technology, Gaithersburg, Maryland 20899}

\author{S.\ -E.\ Park} 

\address{Materials Research Laboratory, The Pennsylvania State
  University, University Park, Pennsylvania 16802 \\ (Present address:
  Fraunhofer-IBMT Technology Center Hialeah, Hialeah, Florida 33010) }

\author{G.\ Shirane} 

\address{Physics Department, Brookhaven National Laboratory,
Upton, New York 11973}

\maketitle

\tightenlines

\begin{abstract} 
  Recent neutron scattering measurements performed on the relaxor
  ferroelectric Pb[(Zn$_{1/3}$Nb$_{2/3}$)$_{0.92}$Ti$_{0.08}$]O$_3$
  (PZN-8\% PT) in its cubic phase at 500~K, have revealed an anomalous
  ridge of inelastic scattering centered $\sim 0.2$~\AA$^{-1}$ from
  the zone center (Gehring {\it et al}., Phys.\ Rev.\ Lett.\ {\bf 84},
  5216 (2000)).  This ridge of scattering resembles a waterfall when
  plotted as a phonon dispersion diagram, and extends vertically from
  the transverse acoustic (TA) branch near 4~meV to the transverse
  optic (TO) branch near 9~meV.  No zone center optic mode was found.
  We report new results from an extensive neutron scattering study of
  pure PZN that exhibits the same waterfall feature.  We are able to
  model the dynamics of the waterfall using a simple coupled-mode
  model that assumes a strongly $q$-dependent optic mode linewidth
  $\Gamma_1(q)$ that increases sharply near 0.2~\AA$^{-1}$ as one
  approaches the zone center.  This model was motivated by the results
  of Burns and Dacol in 1983, who observed the formation of a
  randomly-oriented local polarization in PZN at temperatures far
  above its ferroelectric phase transition temperature.  The dramatic
  increase in $\Gamma_1$ is believed to occur when the wavelength of
  the optic mode becomes comparable to the size of the small polarized
  micro-regions (PMR) associated with this randomly-oriented local
  polarization, with the consequence that longer wavelength optic
  modes cannot propagate and become overdamped.  Below $T_c$ = 410~K,
  the intensity of the waterfall diminishes.  At lowest temperatures
  ($\sim$ 30~K) the waterfall is absent, and we observe the recovery
  of a zone center transverse optic mode near 10.5~meV.
\end{abstract}

\section{Introduction}

Scientific research in the field of relaxor ferroelectrics has surged
markedly over the past several years.  Perhaps the primary driving
force behind the rapidly increasing number of publications in this
area has been the observation of an exceptionally large piezoelectric
response in single crystals of the two complex perovskite systems
Pb[(Mg$_{1/3}$Nb$_{2/3}$)$_{1-x}$Ti$_x$]O$_3$ (PMN-$x$PT) and
Pb[(Zn$_{1/3}$Nb$_{2/3}$)$_{1-x}$Ti$_x$]O$_3$ (PZN-$x$PT) for
compositions $x$ that lie near the morphotropic phase boundary (MPB),
which separates the rhombohedral and tetragonal regions of the phase
diagram.  These systems differ from the classic and well-studied
simple perovskite $ABO_3$ compounds by virtue of the mixed-valence
character of the $B$-site cation.  In the case of PMN and PZN ($x$=0),
the $B$-site is occupied by either Mg$^{2+}$ or Zn$^{2+}$, and
Nb$^{5+}$ ions with a stoichiometry of 1/3 and 2/3, respectively,
which is required to preserve charge neutrality.  This built-in
disorder sharply breaks the translational symmetry of the lattice and
produces a so-called ``diffuse'' phase transition in which the
dielectric permittivity $\epsilon$ exhibits a very large and broad
peak at a characteristic temperature $T_{max}$ that is also strongly
frequency-dependent.  In contrast to the normal ferroelectric parent
compound PbTiO$_3$ (PT), in which the condensation of a transverse
optic (TO) zone center phonon leads to a transition from a cubic
paraelectric phase to a tetragonal long-range ordered state with a
non-vanishing spontaneous polarization below a critical temperature
$T_c = 763$~K, the prototypical relaxor compound PMN exhibits a
diffuse transition at $T_{max} = 230$~K and no spontaneous
polarization at any temperature studied.

Stimulated by the unusual properties of these complex systems, we
performed a recent neutron inelastic scattering study to examine the
lattice dynamics of a high-quality single crystal of PZN-8\%PT,
\cite{Gehring1} the composition for which the piezoelectric response
is largest in the PZN-$x$PT system. \cite{Park1} The PZN-$x$PT phase
diagram as a function of PT concentration $x$ is shown in Fig.~1.  The
maximum piezoelectric activity occurs on the rhombohedral side of the
MPB, which is indicated by the steep dashed line.  Our measurements of
the phonon dispersion of the polar TO mode in the cubic phase at 500~K
revealed an unexpected ridge of scattering centered at a momentum
transfer $q = 0.2$~\AA$^{-1}$, measured from the zone center.  The
ridge extends in energy from $\sim 4$~meV to 9~meV such that, when
plotted as a standard phonon dispersion diagram, it resembles a
waterfall in which the TO branch appears to drop precipitously into
the transverse acoustic (TA) branch.  This anomalous feature,
hereafter referred to as the waterfall, was observed in both of the
two Brillouin zones we surveyed (2,2,$l$), and ($h$,$h$,4), which
cover the two different symmetry directions [001] and [110].  Similar
data taken on a large single crystal of pure PMN at room temperature
also showed the presence of a waterfall, suggesting that this feature
may be common to all relaxor systems. \cite{Gehring2}

%
% ============================= Fig. 1 ============================== %
%
\begin{figure}
 \begin{center}
   \parbox[b]{3.375in}{
     \psfig{file=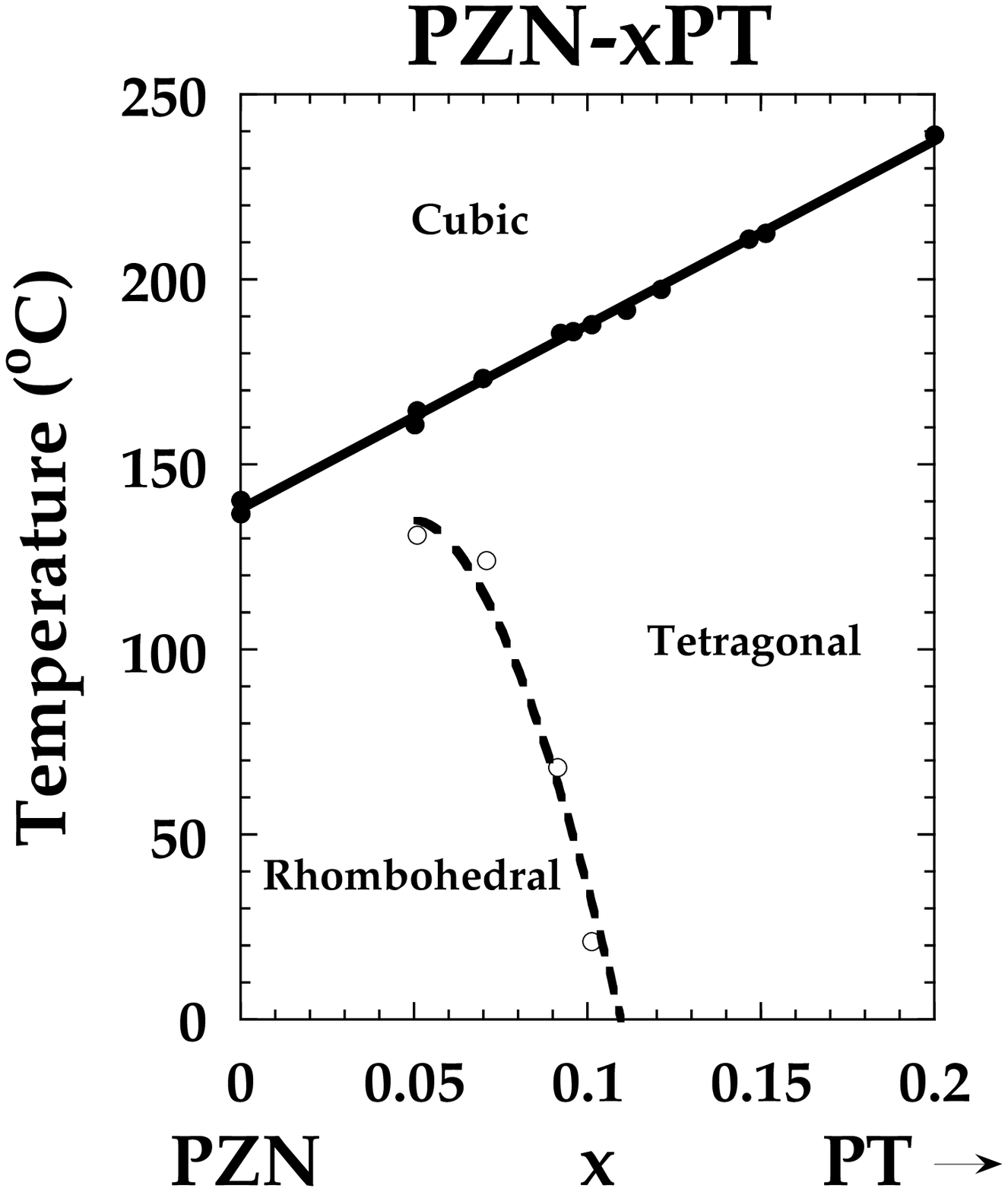,width=2.75in}{\vspace{0.1in} Fig.\ 1.
       \small Phase diagram of the PZN-PT system from Kuwata {\it et
         al}.\ (ref.\ [3]).  The MPB is represented by the dashed line
       between the rhombohedral and tetragonal phases.  An intervening
       monoclinic phase has been discovered in PZT by Noheda {\it et
         al}.\ (ref.\ [4]). }}
 \end{center}
 \label{fig:1}
\end{figure}

The waterfall observed in PZN-8\%PT was speculated to represent direct
microscopic evidence for small, randomly-oriented regions of local
polarization within the crystal that begin to condense at temperatures
far above $T_c = 450$~K.  The existence of these regions of local
polarization was first proposed by Burns and Dacol in 1983 to describe
the disorder inherent to relaxor systems. \cite{Burns} Using
measurements of the optic index of refraction to study ceramic samples
of (Pb$_{1-3x/2}$La$_x$)(Zr$_y$Ti$_{1-y}$)O$_3$ (PLZT), as well as
single crystals of PMN and PZN, \cite{Burns} they discovered that a
randomly-oriented, and non-reversible local polarization $P_d$
develops at a well-defined temperature $T_d$, frequently referred to
as the Burns temperature, several hundred degrees above the apparent
transition temperature $T_c$.  Subsequent studies have provided
additional experimental evidence supporting the existence of $T_d$.
\cite{Mathan,Bokov,Zhao} The spatial extent of these locally polarized
regions was conjectured to be of the order of several unit cells, and
has given rise to the term ``polar micro-regions,'' or PMR.  Because
the size of the PMR is finite, the propagation of long-wavelength
phonons should be inhibited.  This would then lead naturally to a
heavily overdamped TO phonon cross section in the neighborhood of the
zone center.  Indeed, no well-defined TO phonon peaks were found in
either PZN-8\%PT or PMN for $q < 0.2$~\AA$^{-1}$.  Moreover, if one
assumes the waterfall peak position in $q$ gives a measure of the size
of the PMR according to $2\pi / q$, one obtains a value of 30~\AA, or
about 7 -- 8 unit cells, consistent with the conjecture of Burns and
Dacol.

Further evidence in support of our picture correlating the waterfall
with the appearance of the PMR comes from the neutron inelastic
scattering data of Naberezhnov {\it et al}.  As shown in Fig.~2, they
observed a normal TO and TA phonon dispersion in the closely related
relaxor ferroelectric compound PMN at 800~K $> T_d = 617$~K, where no
polar regions remain. \cite{Naberezhnov} Their data were obtained on
the same crystal we studied at room temperature where the waterfall
was observed.  For PZN-8\%PT, the formation of the PMR is estimated to
occur at $T_d \sim 790$~K, which is well above the cubic-to-tetragonal
phase transition at $T_c \sim$ 450~K, and also beyond the temperature
at which the crystal begins to decompose.

%
% ============================= Fig. 2 ============================== %
%
\begin{figure}
 \begin{center}
   \parbox[b]{3.375in}{\psfig{file=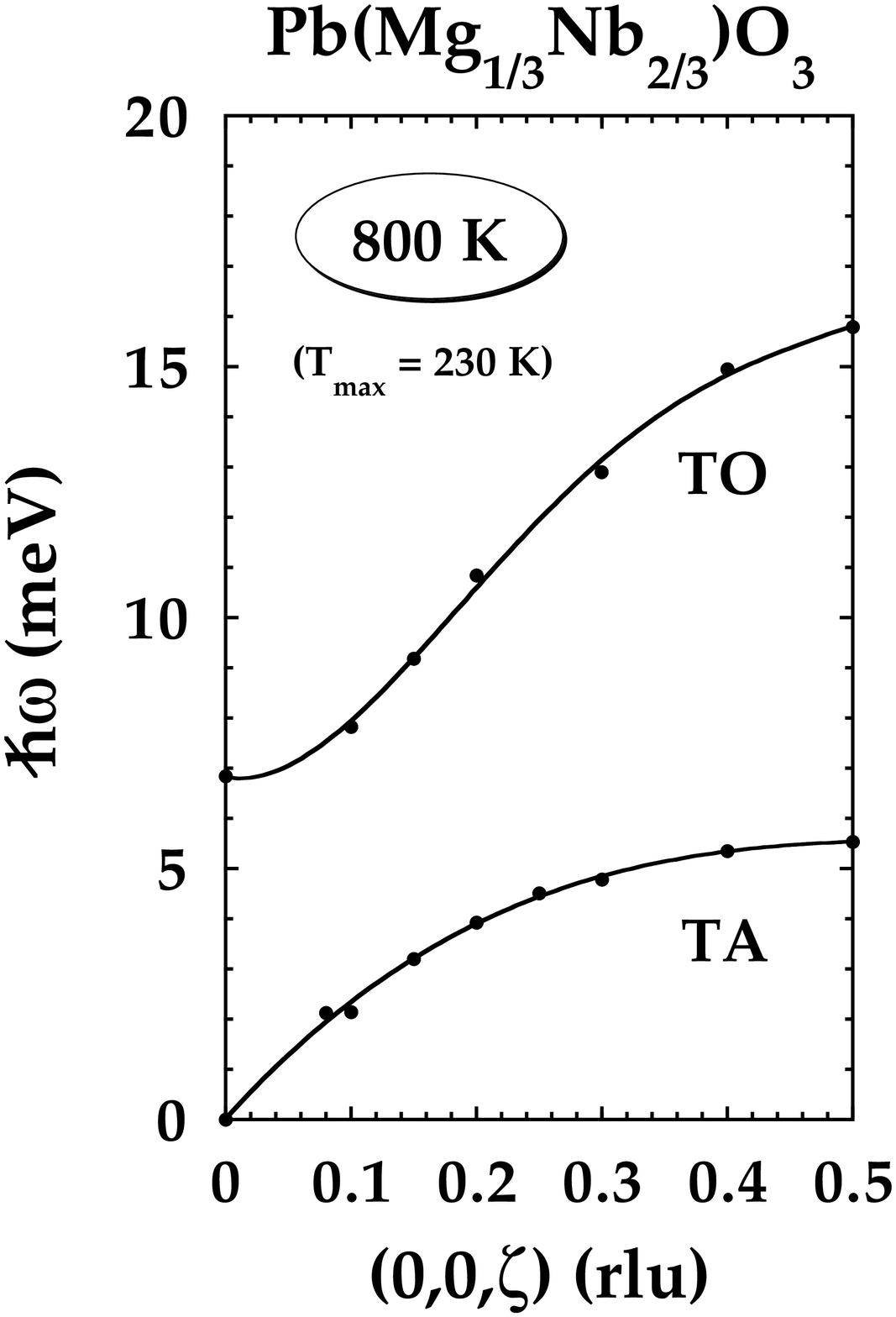,width=2.75in}{\vspace{0.1in}
       Fig.\ 2.  \small Dispersion curves of the transverse acoustic
       (TA) and the lowest-energy transverse optic (TO) modes in PMN
       measured by Naberezhnov {\it et al}.\ (ref.\ [10]) at 800~K,
       far above $T_d$= 617~K. }}
 \end{center}
 \label{fig:2}
\end{figure}

In this paper we present a more complete lattice dynamical study of
pure PZN, in which we also observe an anomalous ridge of scattering
that closely resembles the waterfall seen in PZN-8\%PT and PMN.  PZN
differs from PMN in that it exhibits an explicit structural phase
transformation at 410~K from cubic to tetragonal symmetry.  The value
of $T_{max}$ for PZN is just slightly above $T_c = 410$~K, and thus
much higher than $T_{max} = 230$~K for PMN.  \cite{Park2} We further
present a simple mode-coupling model that clearly relates this unusual
feature of the TO phonon branch (the same branch that goes soft at the
zone center at $T_c$ in PbTiO$_3$) to the PMR.  Our current study does
not, however, cover the critical scattering at small $q$ below 0.075
\AA$^{-1}$.  Extensive studies in this area have already been carried
out using neutron \cite{Vakhrushev-N} and x-ray scattering
\cite{Vakhrushev-X,You} techniques.  As we will show, this is the
portion of the Brillouin zone where the TO mode is overdamped because
of the PMR.

\section{Experimental}

Two single crystals of PZN were used in this study, the first,
labelled PZN\#1, weighs 4.2 grams and the second, labelled PZN\#2,
weighs 2.3 grams.  Both crystals were grown using the high-temperature
flux technique described elsewhere. \cite{Park2} Each crystal was
mounted onto an aluminum holder and oriented with the cubic [001] axis
vertical, thereby giving access to the $(HK0)$ scattering zone.  They
were then placed inside a vacuum furnace capable of reaching
temperatures of up to 900~K.  For temperatures below room temperature,
a specially designed closed-cycle helium refrigerator was used to
cover the range from 25~K to 670~K.

All of the neutron scattering data presented here were obtained on the
BT2 triple-axis spectrometer located at the NIST Center for Neutron
Research.  The (002) reflection of highly-oriented pyrolytic graphite
(HOPG) was used to monochromate and analyze the incident and scattered
neutron beams.  An HOPG transmission filter was used to eliminate
higher-order neutron wavelengths.  The majority of our data were taken
holding the incident neutron energy $E_i$ fixed at 14.7~meV
($\lambda_i = 2.36$~\AA) while varying the final neutron energy $E_f$,
and using horizontal beam collimations of 60$'$-40$'$-S-40$'$-40$'$
and 60$'$-40$'$-S-40$'$-open.

Two basic types of inelastic scans were used to collect data on each
sample.  Constant-$E$ scans were performed by holding the energy
transfer $\hbar \omega = \Delta E = E_i - E_f$ fixed while varying the
momentum transfer $\vec{Q}$.  Constant-$\vec{Q}$ scans were performed
by holding the momentum transfer $\vec{Q} = \vec{k_i} - \vec{k_f}$ ($k
= 2\pi/\lambda$) fixed while varying the energy transfer $\Delta E$.
Using a combination of these scans, the dispersions of both the TA and
the lowest-energy TO phonon modes were mapped out at a temperature of
500~K (still in the cubic phase, but well below the Burns temperature
for PZN of $T_d \sim 750$~K).

It is worthwhile to note that the high temperature data were taken
using neutron energy gain, i.\ e.\ by scanning $E_f$ such that the
energy transfer $\hbar \omega < 0$.  This process is known as phonon
annihilation.  At high temperatures, and not too large energies, the
phonon thermal population factor makes this mode of operation
feasible.  More importantly, the instrumental energy resolution
function contains the factor $\cot 2\theta_a$, where $2\theta_a$ is
the analyzer (HOPG) scattering angle.  Hence the energy resolution
increases with increasing energy because $2\theta_a$ decreases.  The
net result is an increased neutron count rate at the expense of a
broader linewidth, thereby making the relatively weaker transverse
optic mode easier to see.

\section{Inelastic Scattering Data from PZN near {\bf Q}$=(2,0,0)$ }

The waterfall, previously reported in PZN-8\%PT, \cite{Gehring1} is
shown for PZN\#1 at 500~K in Fig.~3.  The data points represent the
peak scattered neutron intensity plotted as a function of $\hbar
\omega$ and $\vec{q}$, where $\vec{q} = \vec{Q} - \vec{G}$ is the
momentum transfer relative to the $\vec{G} = (2,0,0)$ Bragg
reflection, measured along the cubic [010] symmetry direction.  The
solid dots represent data from constant-$\vec{Q}$ scans whereas the
two open circles are data derived from constant-$E$ scans.  These
latter two data points signal the onset of the waterfall regime in
which the TO phonon branch appears to dive into the TA branch, and are
thus connected by a dashed line.  The lowest-energy data points trace
out the TA phonon branch along [010].  Solid lines are drawn through
these points as a guide to the eye, and are nearly identical to those
shown for PMN in Fig.~1.

%The lengths of the vertical
%(horizontal) bars represent the intrinsic phonon FWHM (full width at
%half maximum) linewidths in $\hbar \omega$ ($q$) (i.\ e.\ after
%correcting for the instrumental resolution), and were derived from
%Gaussian least-squares fits to the constant-$\vec{Q}$ (constant-$E$)
%scans.

%
% ============================= Fig. 3 ============================== %
%
\begin{figure}
 \begin{center}
   \parbox[b]{3.375in}{\psfig{file=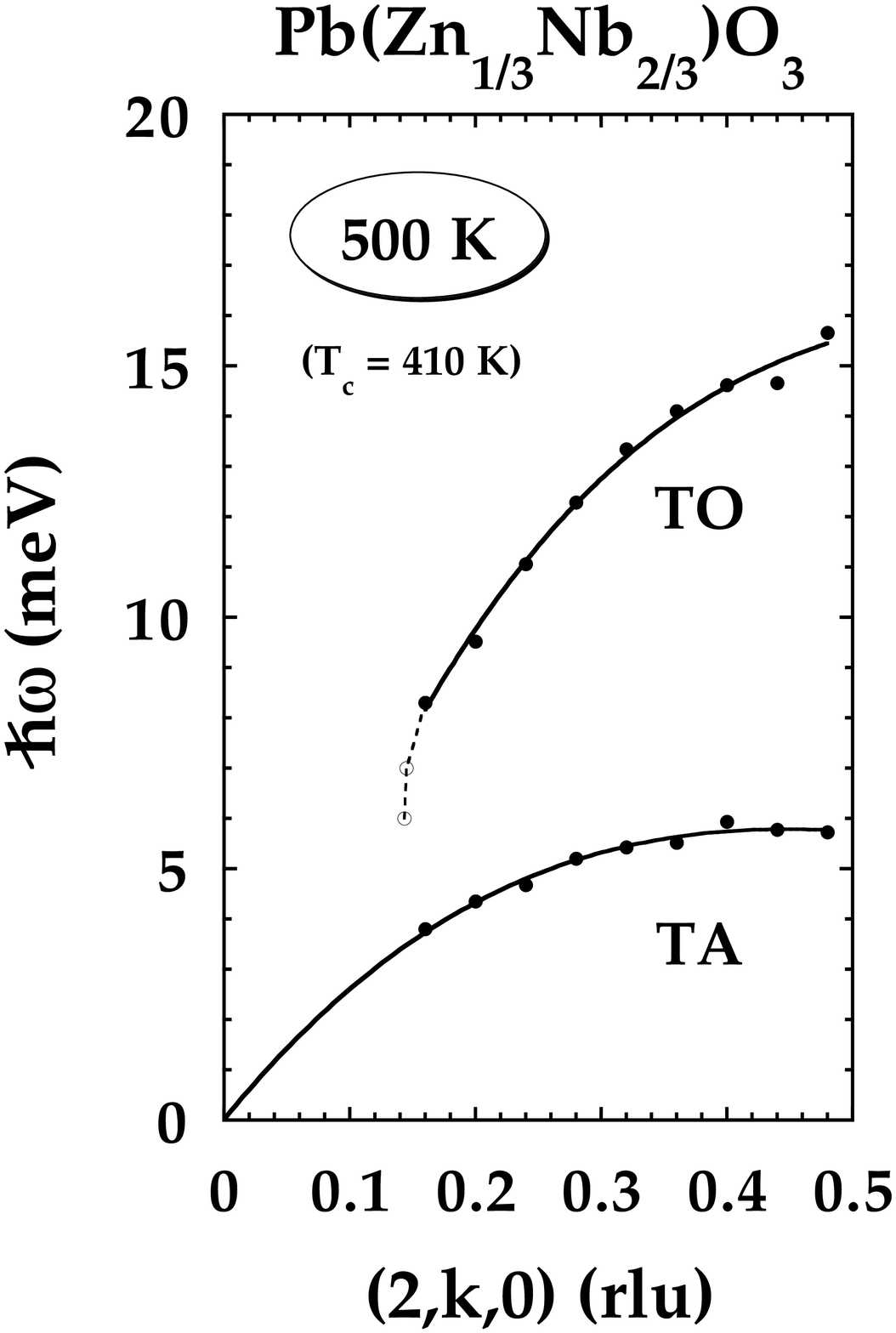,width=2.75in}{\vspace{0.1in}
       Fig.\ 3.  \small Solid (open) dots represent positions of peak
       scattered neutron intensity taken from constant-$\vec{Q}$
       (constant-$E$ scans) at 500~K along the cubic [010] symmetry
       direction in PZN.  Solid lines are guides to the eye indicating
       the TA and TO phonon dispersions.
       }}
% Vertical (horizontal) bars represent the intrinsic phonon
%       FWHM linewidths in $\hbar \omega$ ($q$).
 \end{center}
 \label{fig:3}
\end{figure}

As was the case for PZN-8\%PT measured at the same temperature, the
anomalous ridge of scattering in PZN is centered approximately at 0.14
reciprocal lattice units (rlu) or 0.22~\AA$^{-1}$ (1~rlu$ = 2\pi/a =
1.545$~\AA$^{-1}$).  This is more clearly shown by the series of
constant-$E$ scans plotted in Fig.~4 for $\hbar \omega = $0, -6, -8,
and -12~meV (negative values correspond to phonon annihilation as
described above).  We note that both of the -6 and -8~meV scans are
peaked at the same value of $k$ ($=q$), thereby indicating the
presence of the waterfall.  Remarkably, the 0~meV cross section
exhibits an enormous change in the vicinity of the waterfall
$q$-vector.  This demonstrates that the effects of the PMR extend in
energy all the way from the TO branch through the TA branch and into
the elastic channel.  The -12~meV scan, by contrast, peaks at a larger
momentum transfer $q$ ($\sim 0.24$~rlu) that lies outside the
waterfall regime, and instead represents a genuinely propagating TO
phonon mode.

%
% ============================= Fig. 4 ============================== %
%
\begin{figure}
 \begin{center}
   \parbox[b]{3.375in}{\psfig{file=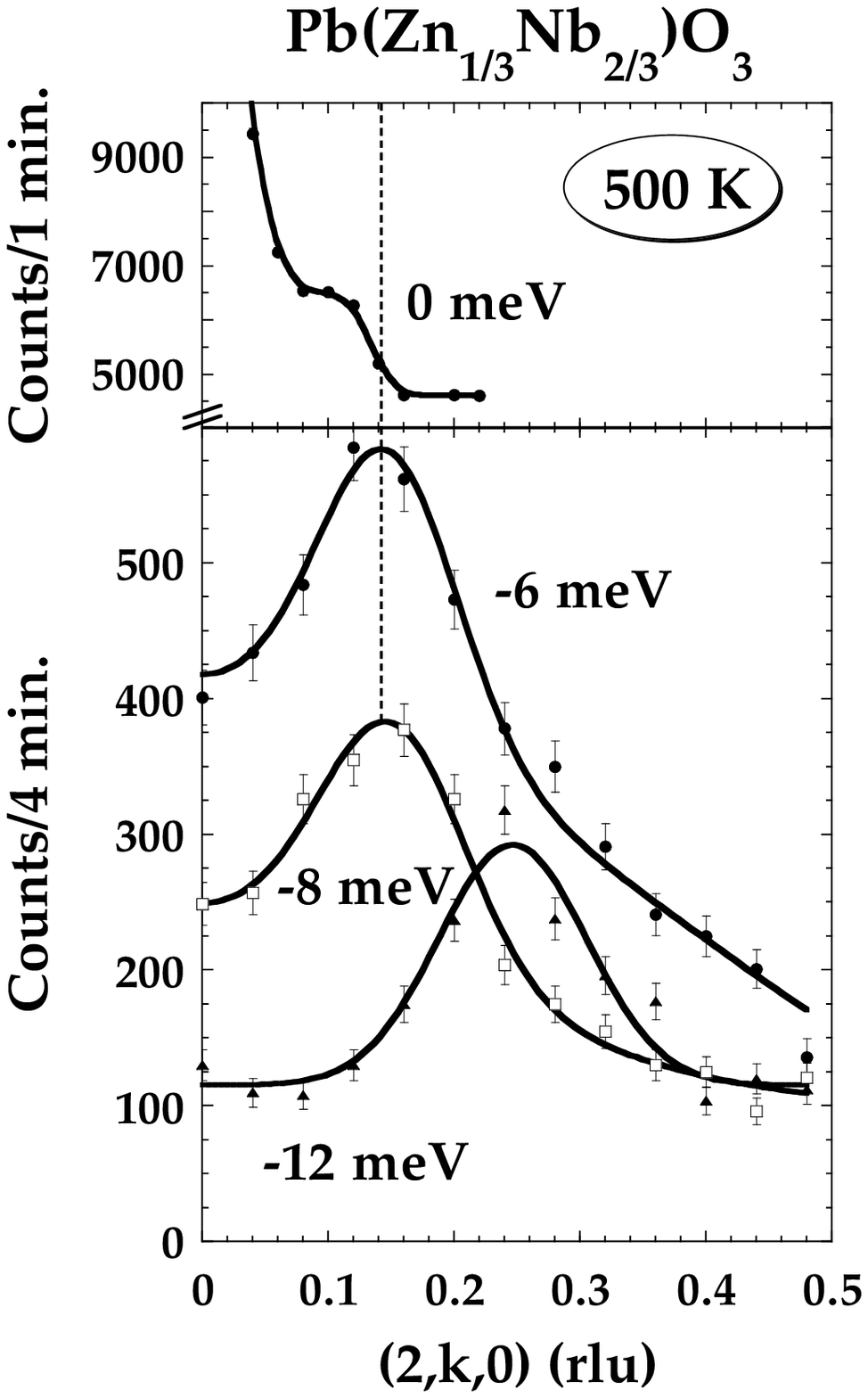,width=2.75in}{\vspace{0.1in}
       Fig.\ 4.  \small Constant-$E$ scans at 0, -6, -8, and -12~meV
       (phonon annihilation) measured at 500~K on PZN crystal \#1.  }}
 \end{center}
 \label{fig:4}
\end{figure}

It is important to realize that peaks in constant-$E$ scans do not
guarantee propagating modes.  This can only be confirmed using
constant-$\vec{Q}$ scans.  Therefore we show four representative
constant-$\vec{Q}$ scans in Fig.~5 for the $q$ values 0.00, 0.16,
0.28, and 0.48~rlu.  For $q > 0.16$~rlu, two distinct peaks are
present, corresponding to the TA and TO phonon modes.  At 0.16~rlu,
one sees the TO mode beginning to become damped while the intensity of
the TA mode has become relatively much more intense.  For $q <
0.16$~rlu, the TA mode was indistinguishable from the elastic peak.
Most striking is the absence of any peak in the constant-$\vec{Q}$
scan for $q$ = 0.00~rlu.  There is no clear peak for the TO mode for
$q < 0.16$~rlu, indicating that these polar optic modes are heavily
overdamped.

%
% ============================= Fig. 5 ============================== %
%
\begin{figure}
 \begin{center}
   \parbox[b]{3.375in}{\psfig{file=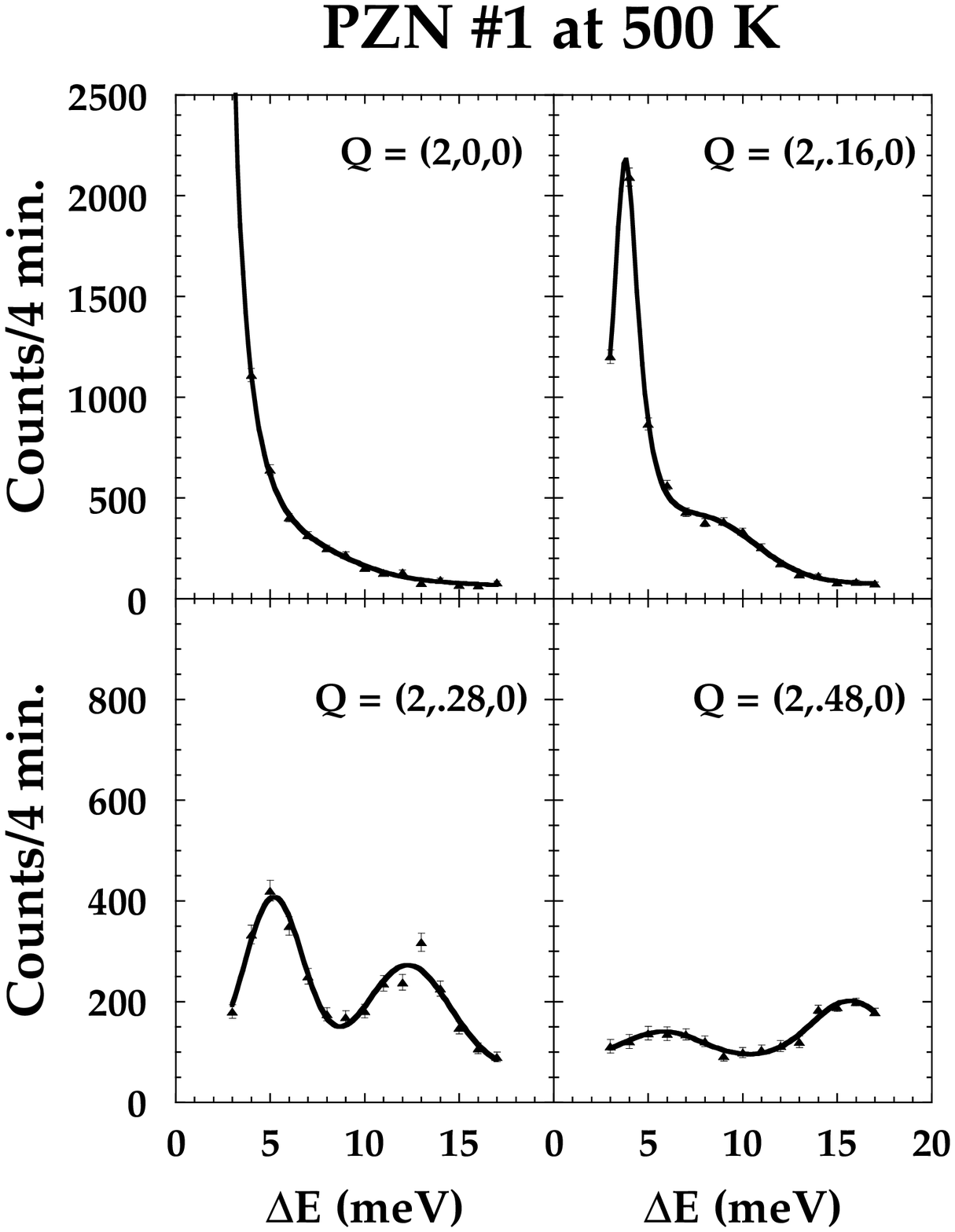,width=2.95in}{\vspace{0.1in}
       Fig.\ 5.  \small Constant-$\vec{Q}$ scans at 0.00, 0.24, 0.28,
       and 0.48~rlu measured at 500~K on PZN crystal \#1. }}
 \end{center}
 \label{fig:5}
\end{figure} 

To illustrate this behavior as clearly as possible, a series of
constant-$\vec{Q}$ scans were collected in steps of 0.04~rlu along
$(2,k,0)$ that spanned the entire Brillouin zone.  After subtraction
of a constant background, the data have been plotted as a contour plot
in Fig.~6.  For ease of comparison, the contour plot has been drawn
with the same vertical and horizontal scales used in Fig.'s 2 and 3.
The intensity data are represented on a logarithmic color scale with
yellow representing the highest intensity, and black the lowest.  At
large $q$ near the zone boundary, the TO and TA phonon branches are
readily apparent as broad green features.  Near the waterfall regime,
the TO branch appears to drop almost vertically into the TA branch
with a corresponding increase in intensity.  Below $q \sim 0.14$~rlu,
the intensity contours show no peak (scanning vertically) that could
correspond to a propagating mode.  The acoustic mode looks to be flat
below 4~meV only because the color scale covers a limited intensity
range above which all data are colored yellow.  The intensity range
was limited on purpose to make the waterfall feature more visible.
Even so, one can notice a very rapid increase in the TA phonon
intensity in the vicinity of the waterfall, suggesting that a transfer
of intensity might be occurring between the TO and TA modes.  To
examine this possibility, we performed model calculations using a
simple mode-coupling scheme to attempt to simulate the anomalous
features reported here in the PZN compound.

%
% ============================= Fig. 6 ============================== %
%
\begin{figure}
 \begin{center}
  \parbox[b]{3.375in}{\psfig{file=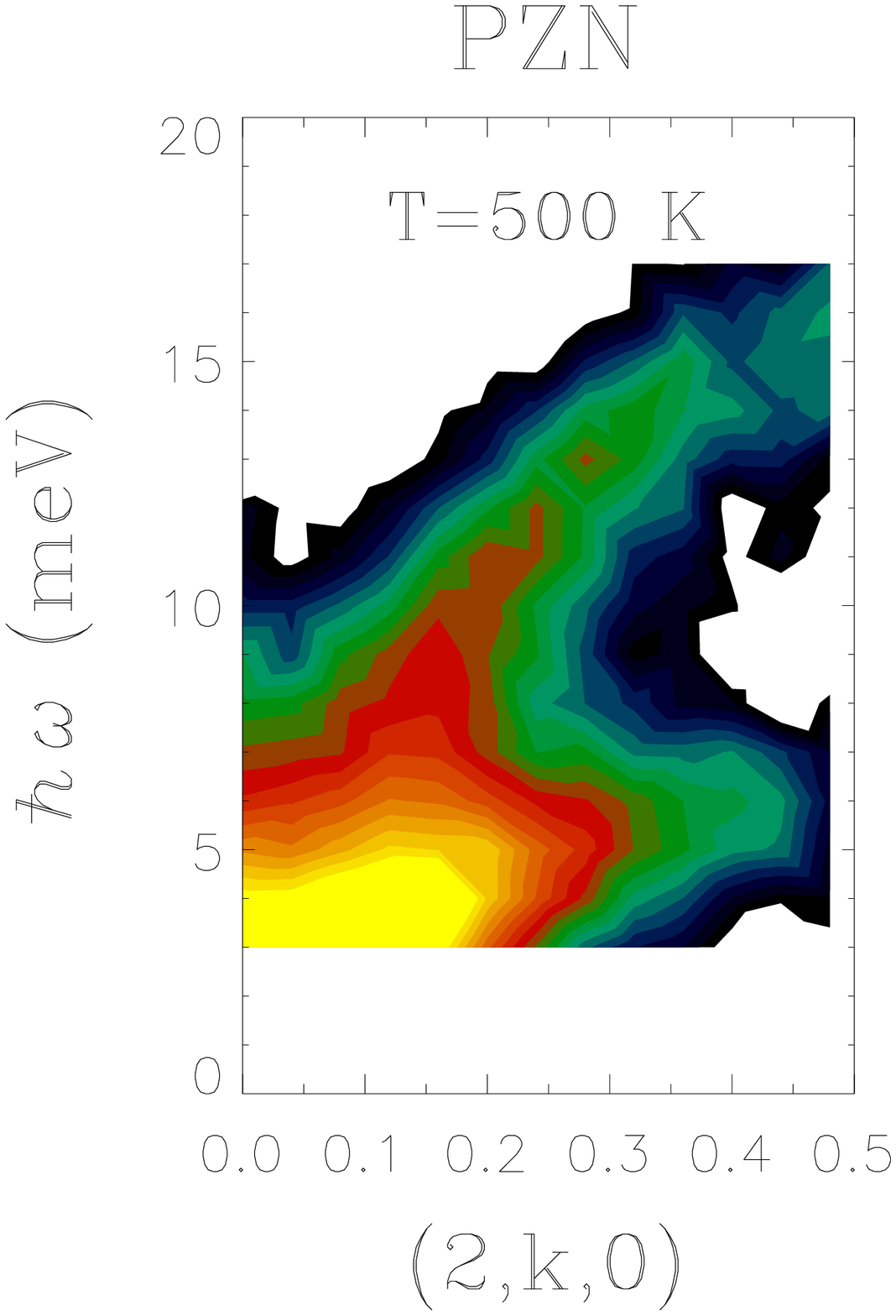,width=2.75in}{\vspace{0.2in}
       Fig.\ 6.  \small Contour map of the background subtracted
       scattering intensity from PZN at 500~K measured near (200).
       The intensity is indicated by a logarithmic color scale that is
       limited to a narrow range in order to better show the
       waterfall.  Yellow is the most intense. }}
 \end{center}
 \label{fig:6}
\end{figure}

\section{Mode-Coupling Model}

During the course of our experiments on relaxor ferroelectrics, a
simple but effective model describing the anomalous low-frequency
lattice dynamics was developed assuming a coupling between the TO and
TA phonon modes.  For the case of neutron energy loss, the scattering
intensity distribution $I$ for two interacting modes with frequencies
$\Omega_1$ and $\Omega_2$, and widths $\Gamma_1$ and $\Gamma_2$, is
given by the expression~\cite{Bullock}

\begin{eqnarray}
I &\sim &[n(\omega) + 1]  \frac{\omega}{A^2 + \omega^2 B^2} \times \nonumber \\ 
  &     &( [( \Omega_2^2 - \omega^2 )B - \Gamma_2 A]F_1^2 + 2\lambda BF_1 F_2 + \nonumber \\
  &     &  [( \Omega_1^2 - \omega^2 )B - \Gamma_1 A]F_2^2 ),
\end{eqnarray}

\noindent
where $A$ and $B$ are given by

\begin{eqnarray}
A &= &(\Omega_1^2 - \omega^2)(\Omega_2^2 - \omega^2) - \omega^2\Gamma_1\Gamma_2, \nonumber \\
B &= &\Gamma_1(\Omega_2^2 - \omega^2) + \Gamma_2(\Omega_1^2 - \omega^2),
\end{eqnarray}

\noindent
and $n(\omega)$ is simply the Bose factor $[e^{(\omega/k_BT)} -
1]^{-1}$.  The quantities $F_{1,2}$ are the structure factors of modes
1 and 2, and $\lambda$ is the coupling strength between the two modes.
Extensive model calculations based on this equation, which has been
shown to describe the behavior of coupled-phonon cross sections quite
well,~\cite{Bullock,Harada} have been performed in collaboration with
K.\ Ohwada at the Institute of Solid State Physics, University of
Tokyo.  \cite{Ohwada}

%
% ============================= Fig. 7 ============================== %
%
\begin{figure}
 \begin{center}
   \parbox[b]{3.375in}{\psfig{file=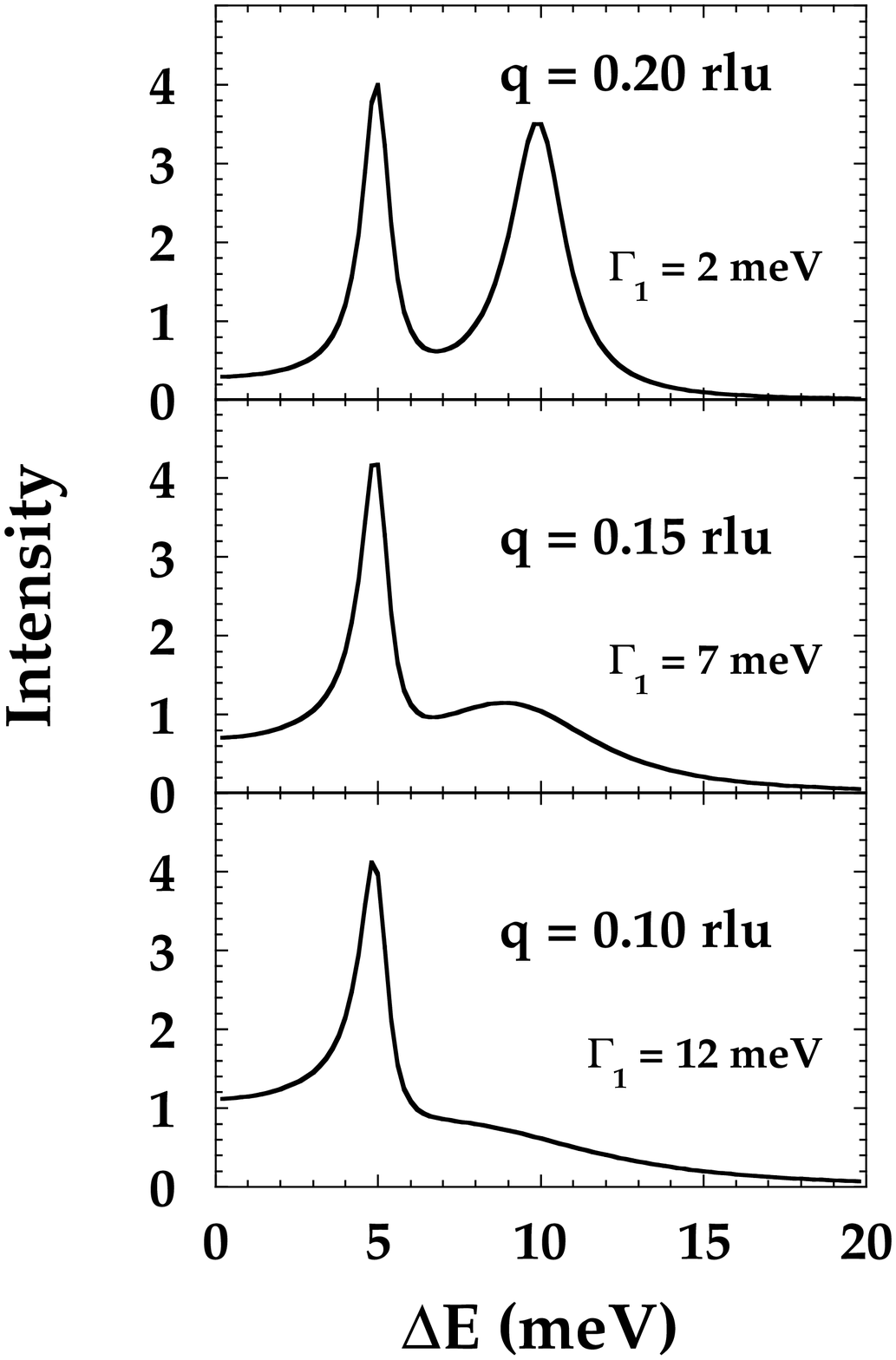,width=2.75in}{\vspace{0.1in}
       Fig.\ 7.  \small Model simulations assuming a coupled-mode
       intensity distribution and a strongly $q$-dependent TO phonon
       linewidth $\Gamma_1$.  Three constant-$\vec{Q}$ scans are shown
       corresponding to $q$ = 0.20, 0.15, and 0.10~rlu, with $q$ =
       0.15~rlu taken to be the reciprocal space position of the
       anomalous waterfall feature. }}
 \end{center}
 \label{fig:7}
\end{figure}

The essential physics behind the mode-coupled model description of the
waterfall is built into the linewidth of the TO mode $\Gamma_1$, which
is assumed to become sharply $q$-dependent as the polar micro-regions
begin to form at the Burn's temperature $T_d$.  If we suppose that the
PMR have an average diameter given by $2\pi/q_{wf}$, where $q_{wf}$
represents the reciprocal space position of the waterfall, then those
optic phonons having $q < q_{wf}$ will not be able to propagate easily
because their wavelength exceeds the average size of the PMR.  These
polar lattice vibrations are effectively impeded by the boundary of
the PMR.  This idea has also been discussed by Tsurumi {\it et al}.
\cite{Tsurumi} The simplest way to simulate this situation is to
assume a sudden and steep increase in $\Gamma_1(q)$ at $q_{wf}$.  For
this purpose, a Fermi-distribution function of $q$ works extremely
well.  Fig.~7 shows several model constant-$\vec{Q}$ simulations based
on this assumption, using the values $q_{wf} = 0.15$~rlu, $\lambda =
10$, $F_1 = 1$, $F_2 = 4$, and $\Gamma_2 = 1$.  For simplicity and
purposes of illustration, the dispersions of both optic and acoustic
modes were ignored by holding the parameters $\Omega_{1,2}$ fixed at
10 and 5~meV, respectively, over the entire Brillouin zone.
Similarly, instrumental resolution effects were not included.

For $q > q_{wf}$ one observes two broad peaks, as expected.  At
momentum transfers $q < q_{wf}$, however, the optic mode becomes
highly overdamped and its profile extends in energy below that of the
acoustic mode.  Alongside each constant-$\vec{Q}$ scan is shown the
corresponding value of $\Gamma_1$ used in the simulation.  The
waterfall thus represents the crossover between a high-$q$ regime, in
which one observes two well-defined peaks corresponding to two
propagating modes, and a low-$q$ regime, in which one observes an
overdamped optic mode plus an acoustic peak.  This simple model cross
section describes all of the experimental observations very well.
Indeed, one can favorably compare the simulated scan at $q$ = 0.15~rlu
in Fig.~7 with the experimental scan for $\vec{Q} = (2,0.16,0)$ shown
in Fig.~5.  One can see now that the waterfall is {\it not} a TO
phonon dispersion at all.  Instead, it is simply a redistribution of
the optic mode profile that is caused by the PMR which force a sudden
change in the optic mode linewidth at a specific $q$ related to the
average size of the PMR.

%
% ============================= Fig. 8 ============================== %
%
\begin{figure}
 \begin{center}
   \parbox[b]{3.375in}{\psfig{file=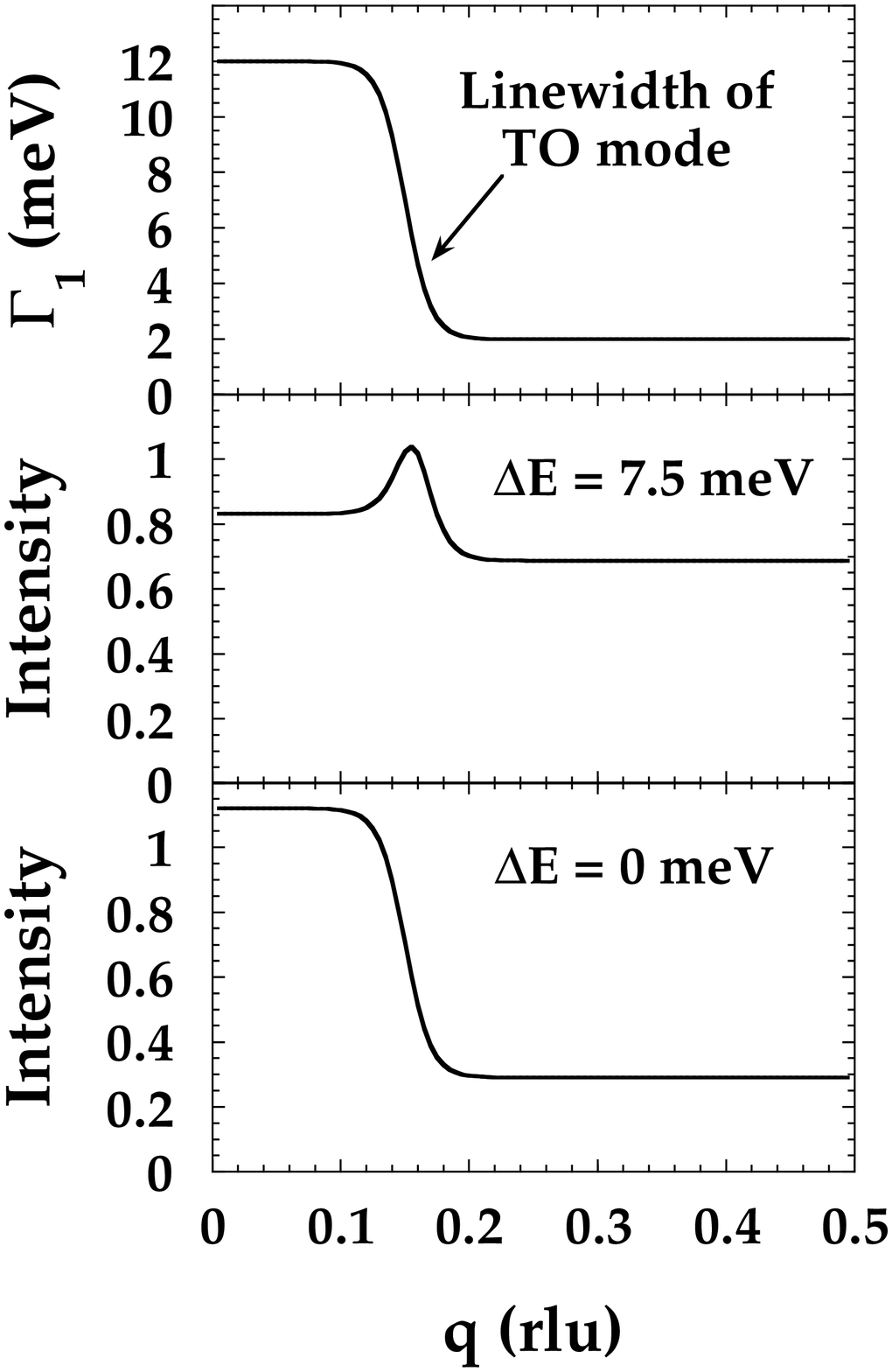,width=2.75in}{\vspace{0.1in}
       Fig.\ 8.  \small Model simulations of two constant-$E$ scans at
       0 and 7.5~meV.  The $q$-dependence of the TO linewidth
       $\Gamma_1$ is shown in the top panel for ease of comparison. }}
 \end{center}
 \label{fig:8}
\end{figure} 

In order to illustrate the basic characteristics of the scattering
cross section within the coupled-mode model, the $q$-dependence of the
optic mode linewidth $\Gamma_1$ is shown in Fig.~8 along with two
simulated constant-$E$ scans at 0 and 7.5~meV.  It is apparent that
the sharp increase in $\Gamma_1$ has a corresponding and pronounced
effect on both cross sections in the vicinity of $q_{wf}$.  The jump
in the elastic cross section occurs because of the sum rule for the
scattering.  Hence the extreme broadening of the TO mode that occurs
near $q_{wf}$ extends underneath the TA branch and into the elastic
channel.  We emphasize that despite the fact that no resolution
effects, Bragg or background terms were included in these
calculations, the agreement with experiment is surprisingly good.
Indeed, the cross section calculated for 0~meV is striking in its
similarity to the experimentally measured elastic cross section shown
in the top panel of Fig.~4, and thus lends strong credence to the
coupled-mode model.

\section{Temperature Dependence}

Our data on the temperature dependence of the waterfall in PZN are
limited by the fact that PZN decomposes at a temperature less than the
Burns temperature.  This prevents us from performing measurements
above $T_d$ which are of interest because we wish to monitor the
recovery of zone center and low-$q$ TO phonons at temperatures where
the polarized micro-regions no longer exist.  PMN, by contrast, has a
lower $T_d$.  Hence Naberezhnov {\it et al}.\ were able to measure the
TO phonon dispersion at 800~K $> T_d$ where they observed a normal TO
phonon branch. \cite{Naberezhnov} Selected phonons were briefly
studied at 860~K in PZN-8\%PT. \cite{Gehring1} There the scattering
intensity associated with the waterfall increased (as expected from
the Bose temperature factor) beyond that measured at 500~K, and the
peak position of the waterfall, measured using constant-$E$ scans,
shifted in towards the zone center, i.\ e.\ from (2,0.14,0) to
(2,0.10,0).  Constant-$\vec{Q}$ data at (2,0.10,0) show a broad peak
around 10~meV.  These data suggest that around 860~K the PMR have
already begun to form inside of a ``uniform,'' or unpolarized crystal.
A complete understanding of the behavior in this temperature range
must await the further study of PMN.

Below the PZN tetragonal-to-rhombohedral phase transition at 410~K the
scattering intensity associated with the waterfall decreases
gradually, and is almost completely gone at 100~K.  This behavior is
summarized in the inset to Fig.~9 where the raw intensity (no
background subtraction) at $\vec{Q} = (2,-.12,0)$ and $\hbar \omega =
7$~meV is plotted versus temperature.  Figure 9 itself shows a
constant-$\vec{Q}$ scan at (2,0,0) and 25~K which clearly shows the
presence of a zone center TO mode.

%
% ============================= Fig. 9 ============================== %
%
\begin{figure}
 \begin{center}
   \parbox[b]{3.375in}{\psfig{file=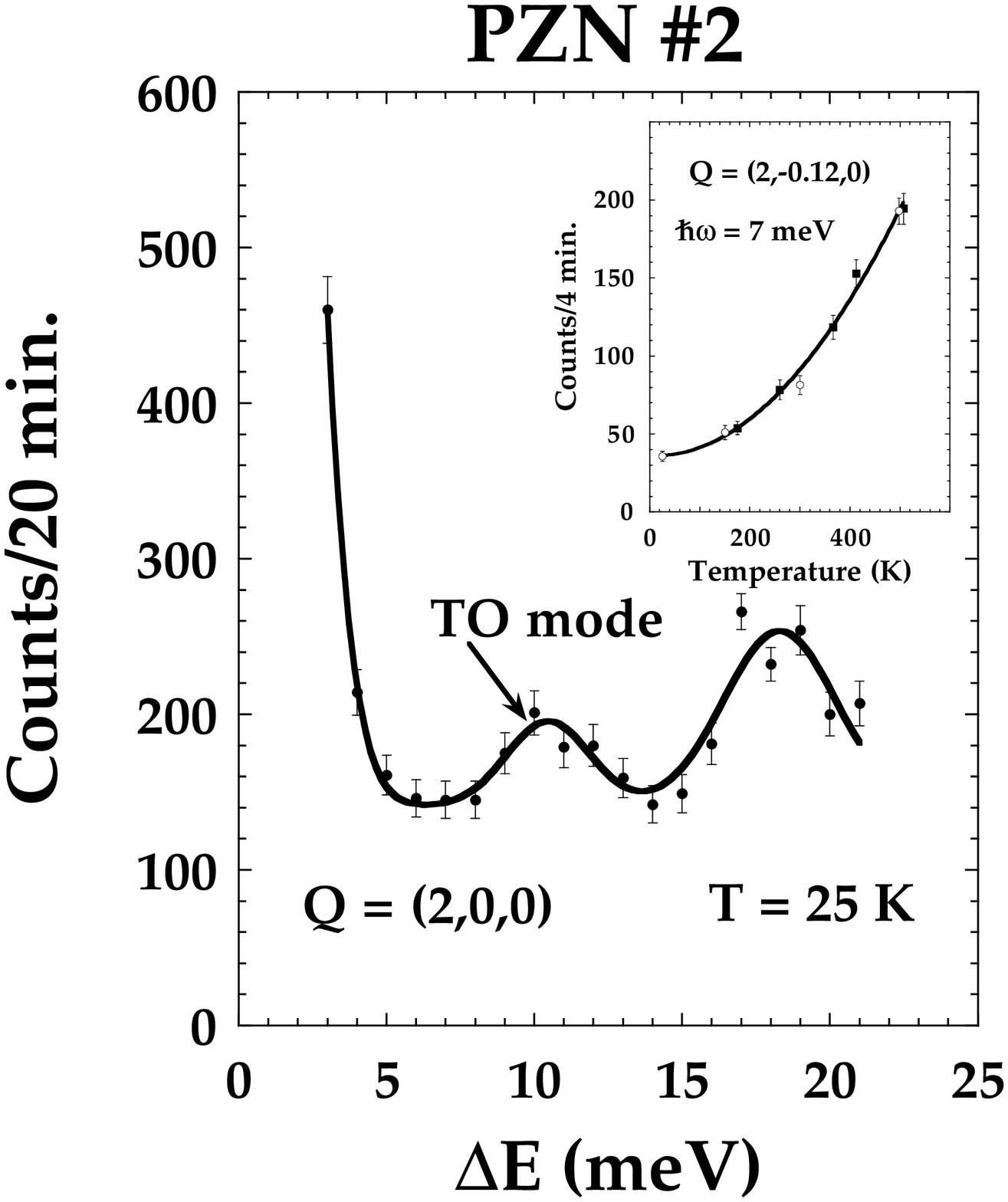,width=2.75in}{\vspace{0.1in}
       Fig.\ 9.  \small Constant-$\vec{Q}$ scan at (2,0,0) and 25~K
       taken on PZN crystal \#2 showing the recovery of the TO zone
       center mode at low temperature. Inset gives the temperature
       dependence of the raw scattering intensity measured at
       $\vec{Q}$ = (2,-0.12,0) at 7~meV (no background subtraction).}}
 \end{center}
 \label{fig:9}
\end{figure}

\section{Discussion}

We have discovered an anomalous ridge of scattering positioned roughly
0.2~\AA$^{-1}$ from the zone center in the relaxor systems PZN-8\%PT,
PZN, and also PMN in their respective cubic phases.  This scattering
was observed in various different Brillouin zones including (220) and
(004) in PZN-8\%PT, and (200), (300), and (220) in PZN.  The ridge of
scattering is extended in energy in such a way as to make the
lowest-lying TO phonon branch appear to drop precipitously into the TA
phonon branch at a specific value of $q$, thereby resembling a
waterfall.  Equally remarkable is the manifestation of a huge elastic
cross section near the same value of $q$.  An unusual feature has also
been observed by Lushnikov {\it et al}.\ for pure PMN using Raman and
neutron inelastic scattering techniques at 77~K over an extended
energy range from 5.2~meV to 6.8~meV, which nearly coincides with that
of the waterfall feature reported here. \cite{Lushnikov} This was
interpreted as evidence of a fracton, i.\ e.\ an excitation on a
fractal lattice.  We have shown that a simple coupled-mode model can
be used effectively to describe our neutron inelastic data by
postulating a sharply $q$-dependent TO phonon linewidth $\Gamma_1$
that exhibits a large step-like increase with decreasing $q$ upon
reaching the waterfall wavevector $q_{wf}$.  In this way we are able
to relate the waterfall directly to the presence of nanometer-sized
polarized micro-regions in the crystal which serve to damp the polar
TO modes at low $q$.  A brief review of the waterfall and related
neutron inelastic scattering measurements on various relaxor systems
has recently been submitted for publication. \cite{Gehring3} It is
hoped that our results with stimulate theoretical calculations of the
precise $q$-dependence of $\Gamma_1$ to allow for comparison with
experiment.  It would be very interesting to investigate the effects
of an applied electric field $E$ on the waterfall, and we have already
begun measurements to this end.

\section{ACKNOWLEDGMENTS}

We would like to thank Y.\ Fujii, K.\ Hirota, B.\ Noheda, K.\ Ohwada,
S.\ B.\ Vakrushev, G.\ Yong, and H.\ D.\ You for stimulating
discussions.  Financial support by the U.\ S.\ Dept.\ of Energy under
contract No.\ DE-AC02-98CH10886, by the Office of Naval Research under
project MURI (N00014-96-1-1173), and under resource for piezoelectric
single crystals (N00014-98-1-0527) is acknowledged.  We also
acknowledge the support of the NIST Center for Neutron Research, U.\ 
S.\ Dept.\ of Commerce, in providing the neutron facilities used in
this work.

\end{document}